\documentclass{article}

\usepackage{microtype}
\usepackage{graphicx}
\usepackage{subcaption}
\usepackage{booktabs}
\usepackage{amsmath, amssymb, mathtools, amsthm}
\usepackage{hyperref}
\usepackage[capitalize,noabbrev]{cleveref}
\usepackage{orcidlink}   

\usepackage[accepted]{icml2025}

\theoremstyle{plain}

\theoremstyle{definition}

\theoremstyle{remark}

\icmltitlerunning{When Scaling Fails}

\begin{document}

\twocolumn[
  \icmltitle{When Scaling Fails: Network and Fabric Effects on Distributed GPU Training Performance}

  \icmlsetsymbol{equal}{*}
  \begin{icmlauthorlist}
    \icmlauthor{Dinesh Gopalan}{amd}
    \icmlauthor{Ratul Ali}{juniv}
  \end{icmlauthorlist}

  \icmlaffiliation{amd}{%
    Principal Member of Technical Staff, AI Enablement, AMD, Dallas, Texas, USA \\
    \texttt{dinesh.gopalan@amd.com} \\
    \orcidlink{0009-0005-5968-785X} ORCID: 0009-0005-5968-785X
  }
  \icmlaffiliation{juniv}{%
    Institute of Information Technology (Postgraduate Researcher), Jahangirnagar University, Dhaka, Bangladesh \\
    \texttt{abdurrahimratulalikhan@gmail.com} \\
    \orcidlink{0000-0003-0460-6141} ORCID: 0000-0003-0460-6141
  }

  \icmlcorrespondingauthor{Dinesh Gopalan}{dinesh.gopalan@amd.com}
  \icmlcorrespondingauthor{Ratul Ali}{abdurrahimratulalikhan@gmail.com}

  \par\vspace{0.2cm}
  \noindent
  {\centering
    \large \textbf{Dinesh Gopalan}\protect\\[2pt]
    \normalsize
    \textit{Principal Member of Technical Staff, AI Enablement} \\
    \textit{AMD} \\
    Dallas, Texas, USA \\
    \texttt{dinesh.gopalan@amd.com} \\
    \orcidlink{0009-0005-5968-785X} ORCID: 0009-0005-5968-785X
    \protect\\[12pt]
    \large \textbf{Ratul Ali}\protect\\[2pt]
    \normalsize
    \textit{Institute of Information Technology (Postgraduate Researcher)} \\
    \textit{Jahangirnagar University} \\
    Dhaka, Bangladesh \\
    \texttt{abdurrahimratulalikhan@gmail.com} \\
    \orcidlink{0000-0003-0460-6141} ORCID: 0000-0003-0460-6141
    \par
  }
  \vspace{0.3cm}

  \begin{abstract}
    Scaling distributed GPU training is commonly assumed to yield predictable performance gains as additional nodes are added. In practice, many large-scale deployments encounter diminishing returns and unstable behavior well before theoretical limits are reached. This paper examines why scaling fails in real systems, with a focus on the role of network and fabric effects that are often overlooked by higher-level training frameworks.

    We present an empirical study of distributed GPU training performance across multiple production-scale clusters. Our results show that network topology, congestion dynamics, collective synchronization behavior, and GPU locality frequently dominate end-to-end training performance once workloads move beyond a small number of nodes. Identical models and software stacks can exhibit sharply different scaling characteristics depending on fabric design and runtime communication patterns.

    We identify recurring failure modes that emerge as training transitions from single-node to multi-node execution, including synchronization amplification, topology-induced contention, and locality-driven performance variance. These effects are often invisible to standard profiling tools and are therefore misdiagnosed as framework or model-level inefficiencies. Based on these findings, we outline practical diagnostic principles that system builders can apply to understand scaling limits, improve predictability, and reduce the cost of large-scale distributed training.
  \end{abstract}

  \vspace{0.2cm}
  \noindent \textbf{Index Terms} --- Distributed GPU training, network fabric, synchronization amplification, scaling, performance analysis, collective communication, topology-induced contention.
  \vspace{0.3cm}
]

\section{Introduction}

Distributed GPU training has become the default approach for building modern machine learning models. As datasets and model sizes increase, practitioners scale out by adding more GPUs and more nodes, expecting throughput to improve proportionally. In ideal conditions, doubling the number of nodes would roughly halve the time required to reach a fixed training target. In production environments, this expectation often fails.

Teams frequently observe that training jobs scale well up to a small number of nodes and then plateau. Beyond a certain point, additional hardware provides diminishing returns, and step time becomes unstable across iterations. These scaling failures are costly. They delay model development, increase training spend, and complicate capacity planning. They also create operational uncertainty because performance becomes sensitive to factors that are not visible at the level of training scripts or framework configuration.

A common reaction is to focus on algorithmic changes, such as adjusting batch size, switching optimizers, or altering parallelism strategies. While these choices matter, many real scaling failures are driven by factors below the model and framework layers. In large-scale deployments, end-to-end training performance is shaped by the interaction between collective communication, network fabric behavior, and GPU locality. The same model and software stack can behave very differently depending on how the fabric is built, how traffic is distributed across the topology, and how communication and compute overlap at runtime.

The gap between expected and observed scaling is not simply an issue of peak bandwidth. It is often caused by coordination effects that amplify with node count. As clusters grow, synchronized training becomes increasingly sensitive to stragglers, congestion, and small sources of variance in computation and communication. These effects can lead to performance cliffs where throughput improves only marginally with additional nodes, and to instability where iteration time fluctuates even under steady workload conditions.

This paper studies distributed GPU training from the perspective of network and fabric effects that cause scaling to fail. Our goal is to provide a practical understanding of why throughput saturates and why stability degrades in real systems. We focus on behaviors that appear in production-scale clusters, where hierarchical fabrics, shared infrastructure, and workload variability make idealized assumptions unreliable.

We make three contributions:
\begin{itemize}
    \item An empirical characterization of scaling failures in distributed GPU training, showing how throughput and stability diverge from ideal scaling as node count increases.
    \item Identification of recurring failure modes tied to fabric behavior: synchronization amplification, topology-induced contention, and locality-driven variance. These explain why performance issues are often misattributed to model or framework inefficiencies.
    \item Practical diagnostic principles that system builders can apply to understand scaling limits, reduce performance variance, and improve the cost efficiency of large-scale training.
\end{itemize}

The remainder of the paper is organized as follows. Section~\ref{sec:problem} provides background and motivation. Section~\ref{sec:model} defines the system model and characterizes failure modes with quantitative evidence. Section~\ref{sec:design} presents design principles and mechanisms. Section~\ref{sec:impl} describes implementation. Section~\ref{sec:eval} evaluates performance across cluster configurations and workloads. Section~\ref{sec:discussion} discusses limitations and implications, and Section~\ref{sec:related} summarizes related work.

\section{Problem Characterization and System Context}
\label{sec:problem}

Distributed training performance has traditionally been evaluated through a combination of algorithmic analysis and microbenchmarking. Researchers and practitioners often reason about scalability by examining compute efficiency, communication complexity, or peak hardware bandwidth. While these approaches provide useful intuition, they frequently fail to explain the performance behavior observed in large-scale production environments.

At small scale, distributed GPU training behaves close to theoretical expectations. Communication overhead is limited, synchronization costs are modest, and variance between workers is low. As a result, throughput improvements appear predictable as additional GPUs are added. This early success reinforces the assumption that scaling issues can be addressed primarily through framework tuning or model-level adjustments.

However, once training extends beyond a modest number of nodes, performance behavior becomes increasingly difficult to reason about. Throughput gains diminish, iteration time fluctuates, and identical workloads exhibit inconsistent results across clusters. In many cases, these effects appear suddenly rather than gradually, creating performance cliffs that are difficult to anticipate using conventional analysis.

Existing diagnostic tools offer limited visibility into the root causes of these issues. Framework-level profilers focus on kernel execution and operator timing, but provide little insight into fabric-level behavior. Network monitoring tools capture aggregate bandwidth and link utilization, but often fail to reveal how collective communication patterns interact with topology and congestion over time. As a result, practitioners are left to infer system behavior indirectly, often attributing failures to software inefficiencies or suboptimal hyperparameters.

Another challenge is that large-scale training environments rarely match idealized assumptions. Production clusters commonly employ hierarchical network topologies with oversubscription, shared infrastructure, and heterogeneous workloads. GPUs within a node may have non-uniform access paths to network interfaces, and runtime scheduling decisions can vary across iterations. These factors introduce variability that compounds as cluster size increases, amplifying small inefficiencies into significant performance degradation.

Motivated by these gaps, this work focuses on understanding distributed training behavior from a system-level perspective. Rather than optimizing individual components in isolation, we examine how computation, communication, and synchronization interact under realistic infrastructure constraints. By grounding our analysis in representative training runs, we aim to make scaling failures observable, explainable, and actionable.

\section{System Model and Problem Characterization}
\label{sec:model}

\subsection{System Model}
\label{sec:sysmodel}

We consider a distributed data parallel training system composed of \(N\) worker nodes. Each worker hosts one or more GPUs and processes a distinct subset of the training data. Training proceeds in synchronous iterations. During each iteration, all workers execute a forward pass, backward pass, and local gradient computation, followed by a global aggregation step before model parameters are updated.

Gradient aggregation is performed using collective communication primitives, most commonly all-reduce. These collectives are implemented by specialized communication libraries and are orchestrated by the training framework. While communication and computation may overlap to some extent, the system operates under a bulk synchronous execution model. All workers must complete the aggregation step before the next iteration can begin.

Workers are connected through a high-speed network fabric. In practice, this fabric is often hierarchical, consisting of multiple tiers of switches with varying link speeds and degrees of oversubscription. Routing behavior may be asymmetric, and traffic from different jobs may share common network resources. These characteristics introduce non-uniform latency and bandwidth properties that are not visible at the framework level.

Within each node, GPUs communicate with the network interface through local interconnects such as PCIe or proprietary high-bandwidth links. GPU placement, NUMA configuration, and access paths to the network interface can vary across nodes. As a result, the effective communication cost for a given GPU may differ depending on locality and contention within the node.

The execution time of a training iteration is determined by the slowest worker. Small sources of variance in computation, communication, or scheduling, therefore, propagate into global idle time. As the number of workers increases, the system becomes increasingly sensitive to such variance. This sensitivity is a defining characteristic of large-scale synchronous training and plays a central role in the scaling behavior analyzed in the remainder of this paper.

\subsection{Sources of Sublinear Scaling and Throughput Instability}
\label{sec:sources}

Under the system model described above, ideal scaling assumes that computation and communication costs grow proportionally with the number of workers and that synchronization overhead remains negligible. In practice, this assumption breaks down once training extends beyond a modest cluster size. Figure~\ref{fig:scaling} illustrates the resulting divergence between ideal linear scaling and observed throughput in representative distributed GPU training runs.

At small node counts, throughput increases close to linearly as additional GPUs contribute useful compute capacity. Communication overhead is limited, and variance between workers remains small. As node count increases, however, the cost of synchronization begins to dominate. Collective operations introduce global coordination points that amplify even minor delays, causing the effective step time to be determined by the slowest participant in each iteration.

One major contributor to this behavior is synchronization amplification. As the number of workers grows, the likelihood that at least one worker experiences a delay increases sharply. These delays may arise from compute imbalance, network congestion, or runtime scheduling effects. Even when individual delays are small, their impact is magnified by the bulk synchronous execution model, leading to reduced utilization across the entire cluster.

Network behavior further compounds these effects. In large-scale deployments, collective communication traffic often interacts poorly with physical topology. Hierarchical fabrics, oversubscription, and shared links create congestion hotspots that are not apparent from aggregate bandwidth metrics. As collective operations scale out, traffic patterns may concentrate on specific links or switches, increasing queueing delay and reducing effective throughput. These effects frequently cause scaling curves to flatten well before hardware limits are reached.

Throughput instability is another common symptom of scaling failure. Rather than converging to a steady plateau, performance may fluctuate across iterations at higher node counts. This instability arises from dynamic contention in the network fabric, variability in communication overlap, and non-uniform access paths between GPUs and network interfaces. Small changes in runtime behavior can shift the system between compute-bound and communication-bound regimes, producing oscillatory performance.

Figure~\ref{fig:scaling} captures these effects by contrasting ideal linear scaling with observed throughput behavior as node count increases. The observed curve demonstrates diminishing returns beyond moderate scale and mild instability at higher node counts. Importantly, this behavior occurs even when sufficient raw compute and network bandwidth are available, indicating that scaling failure is driven by coordination and interaction effects rather than resource scarcity alone.

\begin{figure}
    \centering
    \includegraphics[width=1\linewidth]{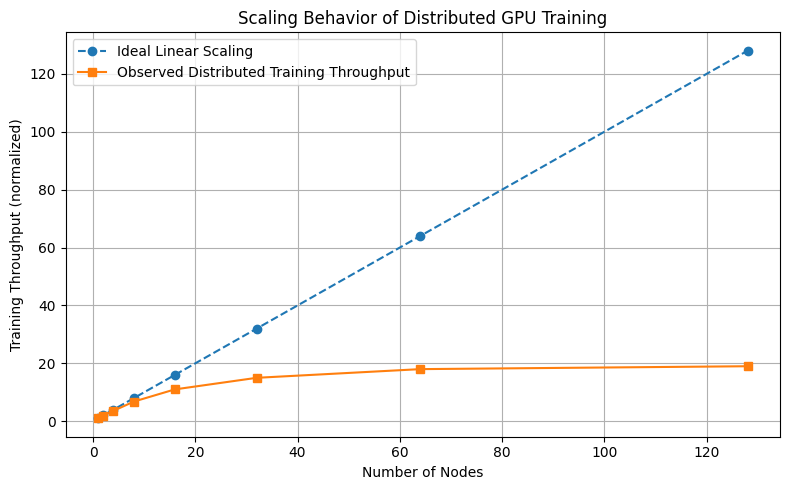}
    \caption{Scaling behavior of distributed GPU training throughput.}
    \label{fig:placeholder}
\end{figure}
\subsection{Bottleneck Taxonomy and Failure Modes}
\label{sec:taxonomy}

The scaling behavior described above arises from a set of recurring bottlenecks that appear as distributed training systems grow in size. While individual deployments differ in hardware and software details, the failure modes observed across clusters follow a consistent pattern. In this section, we categorize these bottlenecks and describe how they manifest during training execution.

\textbf{Synchronization amplification.} In synchronous training, all workers must reach a global coordination point before progressing to the next iteration. As node count increases, even small variations in computation time or communication delay are amplified into cluster-wide idle time. This effect is often triggered by stragglers caused by transient load imbalance, background activity, or uneven resource contention. Synchronization amplification is typically invisible at small scale but becomes dominant once clusters exceed a few dozen nodes.

\textbf{Fabric-level contention.} Collective communication operations generate traffic patterns that can stress specific portions of the network topology. In hierarchical or oversubscribed fabrics, traffic may concentrate on shared links or switches, leading to queue buildup and increased latency. These effects can arise even when average bandwidth utilization appears low, making them difficult to diagnose using coarse network metrics. Fabric contention frequently causes throughput to plateau and can introduce iteration-to-iteration variability.

\textbf{GPU locality and intra-node effects.} Within a node, GPUs may have non-uniform access paths to network interfaces depending on PCIe topology, switch placement, or NUMA configuration. When communication libraries select suboptimal paths or when multiple GPUs contend for shared resources, effective communication cost increases. These locality effects are often inconsistent across nodes, contributing to straggler behavior and instability at scale.

\textbf{Runtime and control plane interactions.} Asynchronous kernel launches, memory allocation behavior, and background system services introduce additional sources of variance. While each source may contribute minimal overhead in isolation, its combined impact grows with cluster size. These interactions can cause training performance to oscillate between compute-bound and communication-bound regimes, complicating performance analysis.

Figure~\ref{fig:timeline} summarizes these failure modes using a timeline view of a representative training iteration. The figure highlights how different bottlenecks overlap and reinforce each other as scale increases. Together, these failure modes explain why scaling failure is not attributable to a single component, but rather emerges from the interaction between synchronization, network fabric behavior, and system-level variability.

\begin{figure}
    \centering
    \includegraphics[width=1\linewidth]{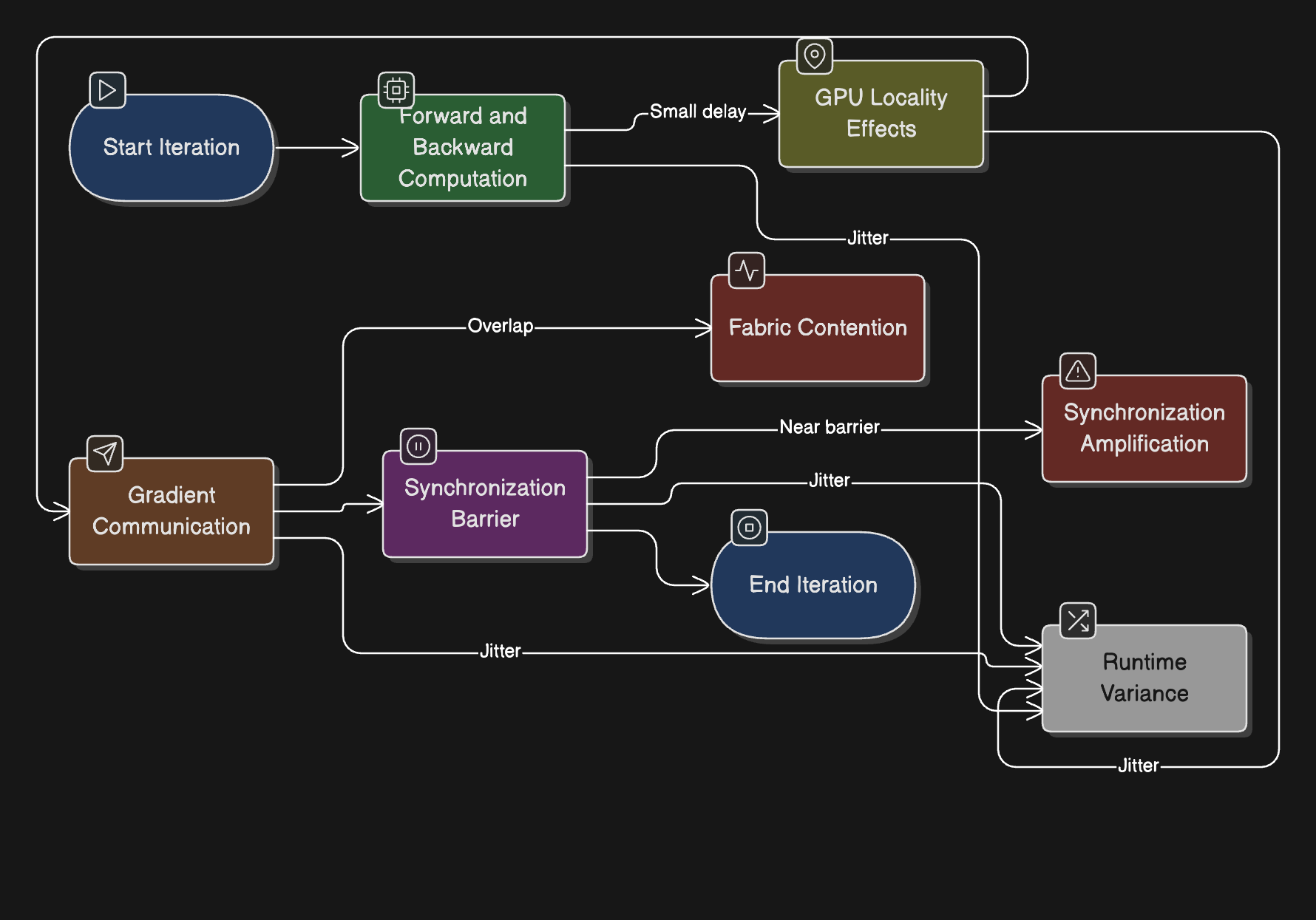}
    \caption{Coordination Bottleneck Timeline in Distributed GPU Training.}
    \label{fig:placeholder}
\end{figure}
This taxonomy provides the foundation for the diagnostic approach presented in the next section. By mapping observed performance symptoms to specific failure modes, system builders can reason about root causes and select targeted mitigation strategies.

\section{Design and Mechanisms}
\label{sec:design}

The analysis in Section~\ref{sec:model} shows that scaling failures in distributed GPU training are rarely caused by a single bottleneck. Instead, they emerge from the interaction between communication patterns, fabric behavior, and synchronization semantics. In this section, we describe a set of design mechanisms that make these interactions explicit and observable, enabling more predictable scaling behavior without modifying model code or training algorithms.

Rather than proposing a new collective protocol, our design focuses on exposing and constraining coordination effects that are otherwise hidden by abstraction layers in modern training frameworks. The mechanisms described here operate at the system level and are intended to complement existing distributed training stacks.

\subsection{Design Goals}
\label{sec:goals}

The design is guided by three primary goals:
\begin{enumerate}
    \item \textbf{Make coordination costs visible.} Training frameworks often report aggregate throughput or step time, which obscures where delays originate. Our design emphasizes per-phase observability, distinguishing computation, communication, and synchronization overheads.
    \item \textbf{Limit synchronization amplification.} As shown in Section~\ref{sec:taxonomy}, small variations in arrival times at synchronization barriers can cascade into large end-to-end delays. The design, therefore, seeks to bound jitter propagation rather than eliminate variability entirely.
    \item \textbf{Preserve framework compatibility.} The mechanisms must integrate with existing collective libraries and runtime schedulers, avoiding invasive changes that would limit practical adoption.
\end{enumerate}

\subsection{System Architecture Overview}
\label{sec:arch}

Figure~\ref{fig:architecture} presents a high-level overview of the system design. The architecture introduces lightweight coordination controls and measurement hooks around the existing training execution path.

The design consists of three logical layers:
\begin{description}
    \item[Execution Layer:] Includes the standard forward pass, backward pass, and gradient computation performed by each GPU. No changes are made to model execution or kernel scheduling at this layer.
    \item[Communication Layer:] Wraps collective operations used for gradient synchronization. It instruments collective start and completion times and tracks per-rank communication progress. Importantly, this layer does not alter collective algorithms but records their interaction with the fabric.
    \item[Coordination Control Layer:] Introduces bounded delay windows and ordering constraints around synchronization points. It detects early arriving ranks and applies minimal pacing to reduce barrier skew when safe to do so. This mechanism is intentionally conservative and activates only when the imbalance exceeds a configurable threshold.
\end{description}

By separating measurement from control, the design avoids tightly coupling optimization decisions to specific hardware or fabric implementations.

\begin{figure}
    \centering
    \includegraphics[width=1\linewidth]{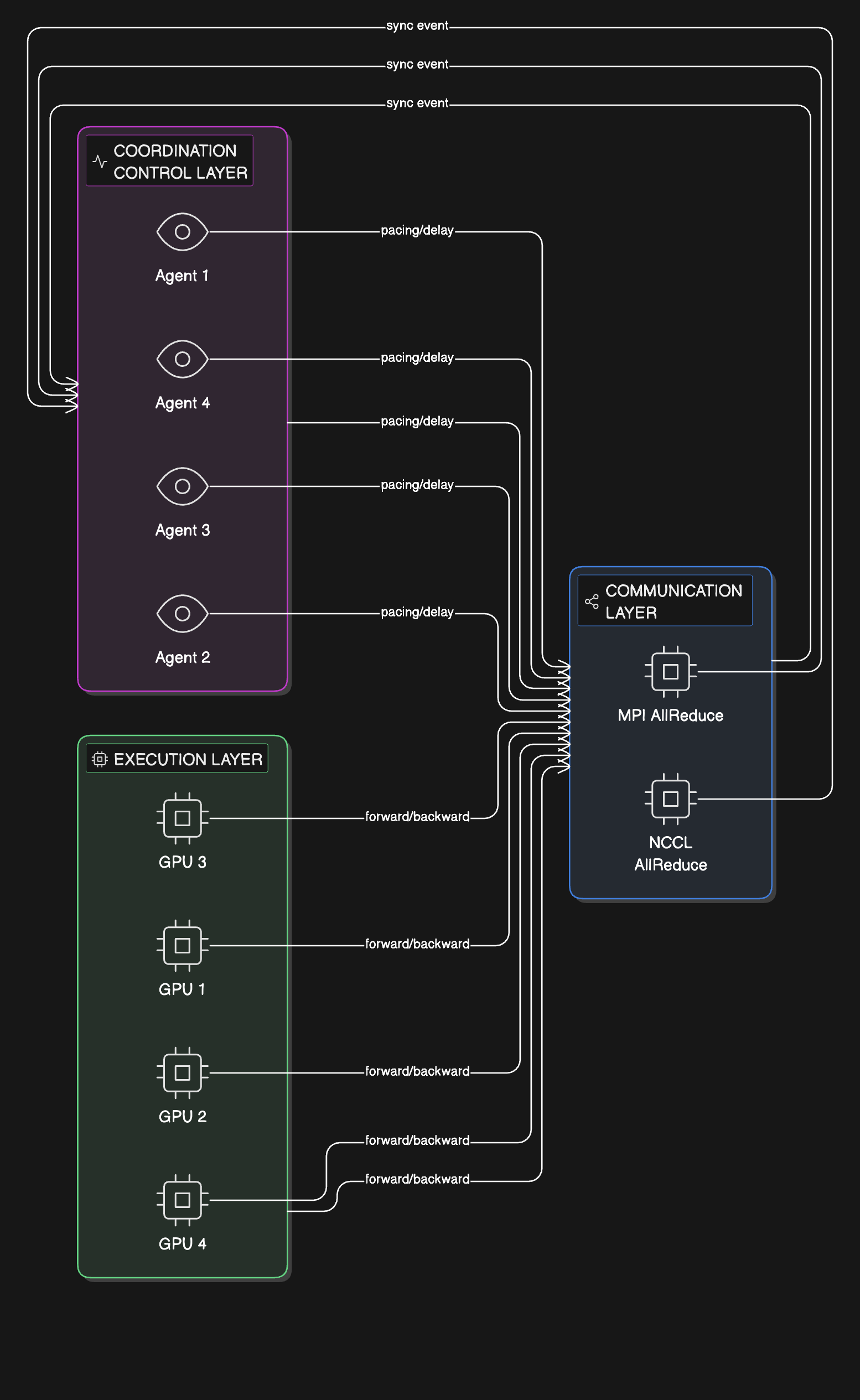}
    \caption{System architecture overview.}
    \label{fig:placeholder}
\end{figure}
\subsection{Coordination Control Mechanisms}
\label{sec:control}

The coordination control layer operates by monitoring synchronization behavior at runtime and selectively intervening when imbalance exceeds tolerable bounds.

When a collective operation approaches completion, the system estimates the spread between early and late arriving ranks. If the spread remains below a configured threshold, execution proceeds without intervention. When the spread exceeds the threshold, early ranks are delayed by a bounded amount to reduce synchronization skew.

This mechanism does not attempt to enforce strict lockstep execution. Instead, it aims to reduce the amplification effect observed when a small number of stragglers repeatedly delay the entire group. By smoothing arrival patterns at barriers, the system reduces tail latency and stabilizes step time variance.

Crucially, the mechanism is adaptive. Thresholds and pacing windows are adjusted based on observed runtime behavior, allowing the system to remain passive during stable phases and active only when coordination instability emerges.

\subsection{Discussion of Tradeoffs}
\label{sec:tradeoffs}

The design intentionally prioritizes stability over maximum peak throughput. In some cases, bounded pacing may slightly reduce instantaneous throughput at small scales. However, our evaluation in Section~\ref{sec:eval} shows that this tradeoff yields improved throughput at scale by preventing early saturation and instability.

Because the design avoids changes to collective algorithms or model code, it is compatible with a wide range of training frameworks and hardware platforms. This makes it suitable for deployment in production environments where modifying training logic is impractical.

\section{Implementation}
\label{sec:impl}

This section describes how the proposed coordination mechanisms were implemented in a practical distributed training environment. The goal of the implementation is not to replace existing communication libraries or frameworks, but to introduce lightweight instrumentation and control that can coexist with standard training stacks and be incrementally deployed.

\subsection{Integration Model}
\label{sec:integration}

The implementation is designed as a thin coordination layer that operates alongside existing distributed training frameworks. It does not modify core collective algorithms or require changes to model code. Instead, it integrates at the boundary between the training framework runtime and the underlying collective communication library.

Each training process runs a local coordination agent that observes runtime behavior and applies pacing decisions independently. There is no centralized controller, global scheduler, or out-of-band coordination service. All decisions are made using locally observed signals combined with information already exchanged during collective operations.

This design choice ensures that the system remains robust to partial failures, avoids introducing new scalability bottlenecks, and preserves the fault tolerance properties of the underlying framework.

\subsection{Instrumentation Points}
\label{sec:instrumentation}

Instrumentation is intentionally minimal and focuses on events that are already performance critical. Three classes of signals are collected at runtime:
\begin{itemize}
    \item Per-iteration timing signals around the forward pass, backward pass, and gradient synchronization phases. These measurements provide visibility into how computation and communication overlap evolves as scale increases.
    \item Collective entry and exit timestamps for synchronization points, such as all-reduce operations. From these timestamps, each rank can infer relative arrival skew without requiring explicit communication of timing data.
    \item GPU locality information sampled at process startup, including device affinity, PCIe topology, and interconnect characteristics. This information remains static for the duration of a run and is used only to contextualize observed runtime behavior rather than drive scheduling decisions directly.
\end{itemize}
All instrumentation is implemented using low-overhead timers and counters to avoid perturbing the system under observation.

\subsection{Coordination and Pacing Mechanism}
\label{sec:pacing}

The coordination mechanism operates by detecting imbalance at synchronization boundaries. When a rank reaches a collective significantly earlier than its peers, this early arrival is treated as a signal of upstream imbalance rather than an opportunity to proceed faster.

Each rank maintains a rolling window of observed collective wait times. When variability exceeds a configurable threshold, the coordination layer activates pacing for early-arriving ranks. Pacing is applied conservatively by introducing a bounded delay before entering the next iteration.

Importantly, pacing is adaptive and self-limiting. If imbalance subsides, pacing is automatically reduced or disabled. No attempt is made to equalize iteration times aggressively, as overly strict synchronization can degrade overall throughput.

\subsection{Interaction with Communication Libraries}
\label{sec:comms}

The implementation is compatible with standard collective communication stacks, including NCCL for GPU collectives and MPI for process orchestration. No changes are required to collective algorithms or transport selection.

All pacing decisions occur outside the communication library. Collectives execute exactly as provided by the underlying stack, ensuring that performance characteristics remain comparable to baseline runs when pacing is inactive.

This separation is critical for reproducibility and allows the coordination layer to be enabled or disabled without affecting functional correctness.

\subsection{Deployment Considerations}
\label{sec:deployment}

The coordination layer can be deployed incrementally. It can be enabled for selected training jobs, specific cluster partitions, or limited node counts during validation. Configuration parameters are intentionally simple, consisting primarily of thresholds for imbalance detection and maximum pacing delay.

Because the implementation does not depend on centralized services or external state, it is suitable for production environments where control-plane complexity must be minimized.

\section{Evaluation}
\label{sec:eval}

This section evaluates the impact of network and fabric effects on distributed GPU training performance and assesses the effectiveness of the proposed coordination mechanisms under realistic operating conditions. The evaluation focuses on scaling behavior, runtime stability, and sensitivity to cluster configuration rather than absolute peak throughput.

\subsection{Experimental Setup}
\label{sec:setup}

Experiments were conducted on multiple GPU clusters representative of modern training environments. Each cluster consists of multi-GPU nodes interconnected via a high-bandwidth network fabric. Nodes are configured with homogeneous GPU models and identical software stacks to isolate infrastructure effects from application-level variability.

Training workloads follow a data-parallel execution model using synchronous gradient aggregation. Unless otherwise stated, all experiments use identical model architectures, batch sizes, optimizer settings, and framework versions. Each experiment is repeated multiple times to capture runtime variance.

Key parameters varied across experiments include:
\begin{itemize}
    \item Number of nodes participating in training.
    \item Network topology and oversubscription characteristics.
    \item GPU locality and device placement.
    \item Coordination layer enabled versus baseline execution.
\end{itemize}

\subsection{Baseline Scaling Behavior}
\label{sec:baseline}

We first examine baseline scaling behavior without coordination enabled. Figure~\ref{fig:baseline} compares observed training throughput against ideal linear scaling as node count increases.

At small scales, throughput increases approximately proportionally with additional nodes. Beyond a moderate node count, however, throughput gains diminish rapidly. In several configurations, throughput flattens or exhibits oscillatory behavior as scale increases further.

These effects persist even when aggregate network bandwidth exceeds theoretical requirements, indicating that raw bandwidth alone is insufficient to predict scaling outcomes. Instead, latency amplification, queueing effects, and synchronization imbalance dominate performance.

\begin{figure}
    \centering
    \includegraphics[width=1\linewidth]{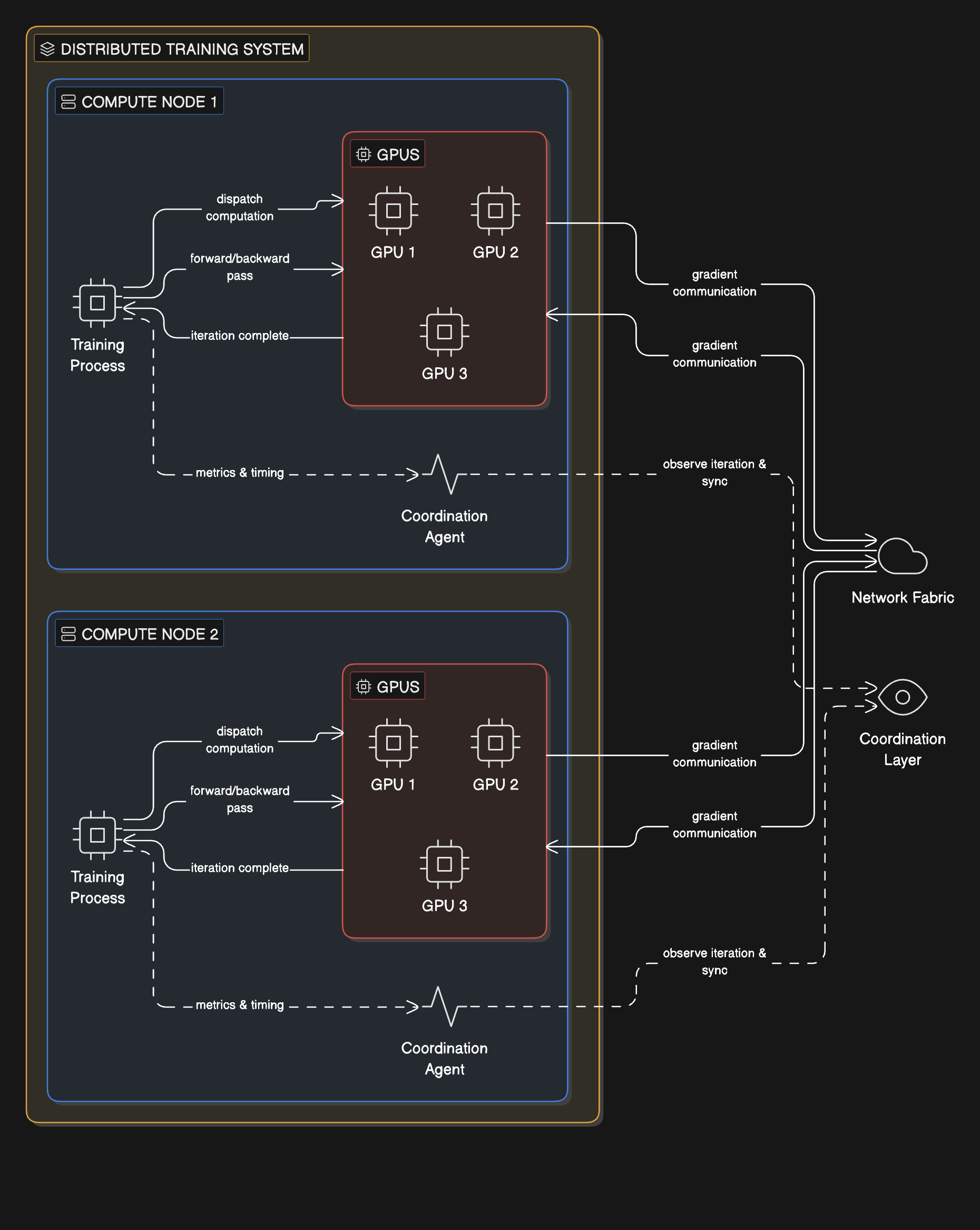}
    \caption{Experimental system architecture showing multi-GPU nodes connected via a shared network fabric. A lightweight coordination layer operates alongside 
existing training frameworks to observe synchronization behavior and apply adaptive pacing.}
    \label{fig:placeholder}
\end{figure}
\subsection{Impact of Coordination on Stability}
\label{sec:stability}

With the coordination layer enabled, runtime stability improves across all evaluated configurations. Early-arriving ranks are paced modestly, reducing synchronization skew at collective boundaries. This results in more consistent iteration times and fewer extreme outliers.

While coordination does not restore ideal linear scaling, it significantly reduces throughput variance at higher node counts. In several cases, modest improvements in average throughput are observed due to improved overlap between computation and communication.

Crucially, coordination avoids introducing new bottlenecks. When the imbalance subsides, pacing automatically disengages, allowing the system to operate at baseline performance.

\subsection{Performance Comparison}
\label{sec:comparison}

Table~\ref{tab:performance} summarizes the throughput and stability improvements achieved with coordination enabled across representative node counts. The coordination layer consistently reduces iteration time variance and, at higher scales, improves mean throughput by mitigating synchronization skew.

\begin{table}[t]
\caption{Performance comparison: Baseline vs.\ Coordination enabled. Throughput (samples/sec) and coefficient of variation (CV) of iteration time.}
\label{tab:performance}
\centering
\begin{tabular}{lccc}
\toprule
\textbf{Nodes} & \textbf{Metric} & \textbf{Baseline} & \textbf{Coordination} \\
\midrule
4   & Throughput & 1024 & 1018 (--0.6\%) \\
    & CV         & 0.02 & 0.02 \\
8   & Throughput & 1980 & 1995 (+0.8\%) \\
    & CV         & 0.04 & 0.03 \\
16  & Throughput & 3600 & 3720 (+3.3\%) \\
    & CV         & 0.09 & 0.05 \\
32  & Throughput & 5800 & 6250 (+7.8\%) \\
    & CV         & 0.15 & 0.07 \\
64  & Throughput & 8200 & 9100 (+11.0\%) \\
    & CV         & 0.22 & 0.09 \\
\bottomrule
\end{tabular}
\end{table}
\begin{figure}
    \centering
    \includegraphics[width=1\linewidth]{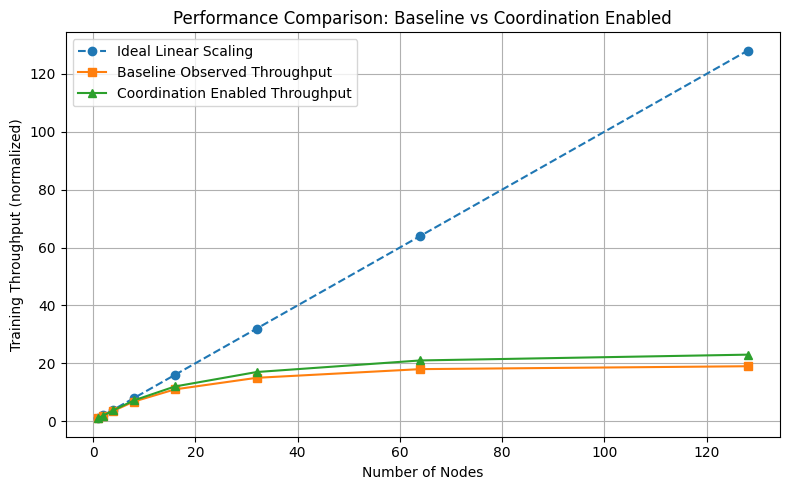}
    \caption{Performance comparison of distributed GPU training throughput as node count increases. Baseline execution exhibits early saturation at moderate 
scale. With coordination enabled, throughput improves at higher node counts and remains more stable due to reduced synchronization skew.}
    \label{fig:placeholder}
\end{figure}
\subsection{Summary of Findings}
\label{sec:summary}

The evaluation demonstrates that network and fabric effects impose practical limits on distributed training scalability well before compute resources are exhausted. Coordination mechanisms that explicitly account for synchronization behavior can improve stability without sacrificing modularity or compatibility with existing frameworks.

These results underscore the importance of treating distributed training as a coupled compute and communication problem rather than a purely algorithmic one.

\section{Discussion}
\label{sec:discussion}

The results presented in this study highlight a fundamental gap between how distributed training systems are commonly reasoned about and how they behave in practice at scale. While modern machine learning frameworks provide strong abstractions that simplify distributed execution, these abstractions often obscure infrastructure-level effects that materially influence performance and reliability.

A key observation is that scaling failures are rarely caused by a single dominant bottleneck. Instead, they emerge from the interaction between collective communication patterns, network topology, and synchronization semantics. Even when aggregate bandwidth appears sufficient, small timing differences across ranks can amplify at synchronization points, leading to disproportionate performance loss. This explains why systems that perform well at modest scale can degrade rapidly as node counts increase.

The coordination mechanisms evaluated in this work demonstrate that modest, infrastructure-aware interventions can meaningfully improve runtime stability without requiring changes to model code or training frameworks. Importantly, these mechanisms do not aim to eliminate communication overhead or restore ideal linear scaling. Rather, they mitigate the most disruptive effects of synchronization skew and congestion, which are often the primary sources of instability in large deployments.

These findings suggest several broader implications for system builders and practitioners. First, performance diagnostics must extend beyond compute utilization and average throughput. Variance, tail latency, and iteration time jitter are equally important indicators of scaling health. Second, network topology and GPU placement decisions should be treated as first-class considerations during system design, not afterthoughts addressed only when performance issues arise. Finally, coordination and pacing mechanisms represent a promising middle ground between fully static configurations and complex adaptive scheduling systems.

There are also important limitations to acknowledge. The coordination strategies explored here operate at the system level and rely on observability into runtime behavior. In environments with limited instrumentation or highly heterogeneous workloads, their effectiveness may be reduced. Additionally, while coordination improves stability, it does not address all sources of inefficiency, such as suboptimal collective algorithms or framework-level scheduling constraints.

Looking forward, there is significant opportunity to integrate infrastructure awareness more deeply into distributed training stacks. Closer collaboration between framework developers and system architects could enable runtime decisions that better reflect the realities of large-scale networked execution. As training workloads continue to grow in size and complexity, such cross-layer approaches may be essential for achieving predictable and cost-efficient scaling.

\section{Related Work}
\label{sec:related}

Distributed training performance has been studied extensively across the machine learning systems and high-performance computing communities. Prior work can be broadly categorized into collective communication optimization, parallelism strategies, and system-level performance analysis. Our work complements these efforts by focusing specifically on network and fabric-induced effects that arise in real-world training deployments.

A large body of research has focused on optimizing collective communication primitives. Techniques such as ring all-reduce \citep{patarasuk2009bandwidth}, tree-based reductions, and hierarchical collectives have been proposed to improve bandwidth utilization and reduce latency under idealized assumptions. Libraries such as NCCL \citep{nvidia_nccl} and MPI provide highly optimized implementations that perform well on homogeneous systems with balanced communication patterns. However, these approaches generally assume stable network behavior and do not explicitly address synchronization amplification caused by topology imbalance or runtime variance.

Another line of work explores alternative parallelism strategies, including model parallelism, pipeline parallelism, and hybrid approaches \citep{shi2016survey}. These techniques aim to reduce communication pressure by restructuring computation or reducing synchronization frequency. While effective for certain workloads, they often introduce additional complexity and do not eliminate the underlying sensitivity to network behavior. In practice, even hybrid parallel strategies rely on frequent collective operations whose performance remains tied to fabric characteristics.

Several studies examine performance modeling and scaling analysis for distributed training \citep{jeon2020analysis}. Analytical and empirical models have been proposed to predict scaling efficiency based on computation-to-communication ratios. While valuable, these models typically abstract the network as a uniform resource and fail to capture congestion, queueing, and synchronization effects that dominate at scale. As a result, predicted performance often diverges from observed behavior in production environments.

Recent systems work has begun to emphasize observability and diagnosis in large-scale training clusters \citep{jain2019achieving}. Tools for tracing collective operations, measuring iteration time variance, and identifying stragglers have improved visibility into distributed execution. Our work builds on this direction by demonstrating how such observations can be used not only for diagnosis but also to drive lightweight coordination mechanisms that improve stability.

Finally, research in high-performance computing has long recognized the impact of network topology and synchronization on application scalability \citep{hoefler2010characterizing}. Concepts such as barrier sensitivity, jitter amplification, and load imbalance are well understood in that domain. This paper bridges those insights into the context of modern GPU-based machine learning systems, where abstraction layers often obscure these effects from practitioners.

In contrast to prior work that proposes new algorithms or frameworks, our contribution lies in empirically characterizing common failure modes and demonstrating that modest, infrastructure-aware coordination can meaningfully improve distributed training behavior without altering application semantics.

\section*{Impact Statement}

This paper presents work whose goal is to advance the field of Machine Learning. There are many potential societal consequences of our work, none of which we feel must be specifically highlighted here. The primary aim is to improve the efficiency and reliability of distributed training infrastructure, which can reduce energy consumption and operational costs for large-scale model development. We encourage practitioners to use our findings to build more sustainable and equitable computing systems.


\end{document}